# Cascaded Complementary Filter Architecture for Sensor Fusion in Attitude Estimation

by Parag Narkhede [1], Shashi Poddar [2], Rahee Walambe [3], George Ghinea [4,*], and Ketan Kotecha [3]

[1] Symbiosis Institute of Technology, Symbiosis International (Deemed University), Pune 412115, India
[2] CSIR-Central Scientific Instruments Organisation, Chandigarh 160030, India
[3] Symbiosis Centre for Applied Artificial Intelligence, Symbiosis International (Deemed University), Pune 412115, India
[4] Department of Computer Science, College of Engineering, Brunel University, London UB8 3PH, UK
[*] Author to whom correspondence should be addressed.

**Abstract**

Attitude estimation is the process of computing the orientation angles of an object with respect to a fixed frame of reference. Gyroscope, accelerometer, and magnetometer are some of the fundamental sensors used in attitude estimation. The orientation angles computed from these sensors are combined using the sensor fusion methodologies to obtain accurate estimates. The complementary filter is one of the widely adopted techniques whose performance is highly dependent on the appropriate selection of its gain parameters. This paper presents a novel cascaded architecture of the complementary filter that employs a nonlinear and linear version of the complementary filter within one framework. The nonlinear version is used to correct the gyroscope bias, while the linear version estimates the attitude angle. The significant advantage of the proposed architecture is its independence of the filter parameters, thereby avoiding tuning the filter's gain parameters. The proposed architecture does not require any mathematical modeling of the system and is computationally inexpensive. The proposed methodology is applied to the real-world datasets, and the estimation results were found to be promising compared to the other state-of-the-art algorithms.

## 1. Introduction

Micro-electro-mechanical systems (MEMS)-based attitude estimation is an active area of research in navigation systems. It is the art of computing the orientation of an object in three-dimensional space. Accurate orientation estimation plays a critical role in aerospace and robotics application, unmanned vehicle navigation, health care applications, safety devices of older people, etc. [1]. Different modalities, such as inertial sensors, LIDARs and cameras, have been used in attitude-estimation applications. Amongst these, inertial sensors are the most popular sensors used for attitude estimation.

Sensor fusion is the process of combining information from two or more sensors to obtain improved accuracy and specific inferences that could not be possible using a single sensor alone [2]. The use of multiple sensors helps with overall performance improvement, increases temporal and special coverages, and adds to the robustness of the system [3]. Inertial sensors (gyroscope and accelerometer) and the magnetometer are used to estimate attitude in all three axes. When integrated to obtain the orientation angle, the angular rates from the gyroscope start drifting over time. This restricts the gyroscope to use as a standalone measurement unit for attitude estimation. A tri-axial accelerometer provides additional information regarding the roll and pitch angles of the $x$ and $y$ axis, respectively. Since it measures the acceleration in terms of the earth's gravity, the axis pointing towards the earth's center cannot observe the change in its measurements, and hence the yaw angle cannot be estimated using an accelerometer. A magnetometer measures the strength of the magnetic field present in its vicinity, and is used to estimate Yaw angle. Tseng et al. [4] simulated the dynamic responses of inertial sensors and showed that the gyroscope possesses a better high-frequency response, whereas the accelerometer and magnetometer have a good low-frequency response. This complementary nature of these sensors makes them the right choice for sensor fusion applications.

Several techniques have been reported in the literature for performing sensor fusion, including the Kalman filter (KF) and its variants, like the extended Kalman filter (EKF), particle filter, unscented Kalman filter (UKF), complementary filter and its variants, etc. A nonlinear version of KF, i.e., EKF, is a widely adopted attitude estimation technique and is still popular amongst researchers [5,6]. Active research is ongoing to incorporate techniques like machine learning [7], statistical methods [8] and fuzzy logic [9,10] to improve the estimation accuracy of KF. However, KF requires the system's mathematical model to be known correctly and is dependent on the system noise parameters. Additionally, KF involves complex inverse matrix operations, increasing its computational complexity [11]. Although several advancements have been made in KF, they are still computationally complex in nature [12]. The stochastic techniques are generally computationally intensive, requiring parameter tuning, and also suffer from divergence due to numerical errors [13]. Alternatively, the deterministic approaches to obtaining the system estimates involve an iterative process and require more computational time.

In order to overcome some of the issues with KF, a complementary filter (CF) has been developed by researchers. CF does not require any knowledge of the system environment or the complex system model [12]. The complementary filter has evolved from a linear to non-linear version, as the linear CF (LCF) was unable to estimate the bias online, leading to inaccurate estimation [14]. The non-linear CF (NCF) is based on the proportional-integral controller, in which the proportional part manages the frequency changeover between two sensors and the integral part handles the gyroscope bias. Mahony et al. [15], in 2008, proposed a CF version in a special orthogonal group and Madgwick et al. [16], in 2011, proposed a gradient-descent-based CF for attitude estimation. These algorithms gained huge popularity amongst researchers owing to their robustness and accuracy. Santos et al. [17] demonstrated that, in quadcopters, the non-linear techniques work better compared to the linear ones. Wu et al. [12] developed a computationally lighter and gradient-descent-based framework of linear complementary filter for attitude estimation. Non-linear CF is used to fuse inertial sensor measurements with the camera [18], depicting its prospects for other applications. A four-parameter-based hybrid complementary filter was proposed by Young in 2020 [19] for attitude estimation application, and is a computationally inexpensive version of Madgwick's filter.

While incorporating CF in an application, the gain parameters of the CF need to be appropriately tuned. The manual selection of these parameters is popular amongst researchers. Fuzzy adaptive versions of the CF are also employed widely for accurate estimations [20,21]. Silva et al. [22] proposed a Kalman-filter-like methodology by considering the system noise characteristics for CF parameter tuning. Some of our earlier works aimed to estimate these gain parameters automatically using optimization techniques [23] and a probabilistic multiple-model-based approach [24].

The deviation observed in the measured magnetic field and gravity vector was considered for tuning CF parameters by Yi et al. [25]. Although multiple adaptive techniques have been developed, they still require prior knowledge of the range of gain parameters.

Alternatively, Foxlin et al. [26] incorporated the Kalman filter and the complementary filter in a unified structure known as the Complementary- Kalman filter (CKF). Like KF, error models were considered while designing CKF, and a feedback mechanism is employed for estimating error. In CKF, a Kalman filter is used to estimate the gyroscope bias, and the complementary filter is then used for attitude estimation. Zhang and Reindl proposed a complementary separate-bias Kalman Filter to determine pedestrian motions [27]. Gyroscope and camera measurements are fused using CKF in [28], wherein a fuzzy adaptive mechanism provides robustness to the filter against varying system dynamics. Li et al. [29] developed a CKF-based indoor navigation system, where the errors in position, velocity, and direction are tracked using the fusion of ultra-wideband sensor and IMU. Yang and Sun [30] proposed a fuzzy-logic-based adaptive CKF technique for the accurate and safe landing of the UAVs. Although CKF was claimed to be a robust estimator, the presence of the KF adds to the computational complexity of the algorithm and involves multiple-matrix inverse operations. Therefore, here, the substitution of KF in CKF with a non-linear complementary filter is proposed, as well as applying the linear complementary filter for attitude estimation. This novel technique of cascaded complementary filter is inspired by the architecture of CKF, and is experimented on the attitude estimation task. Although NCF and LCF's performance depends on their parameters, the proposed cascaded structure does not require any parameter tuning, which is generally manual, tedious, and time-consuming. This means the proposed cascaded structure has a lower computation cost. The contributions of this paper are as follows:

1. A novel architecture of cascaded complementary filter for attitude estimation;
2. The proposed cascaded complementary filter does not require any specific parameter tuning;
3. It is computationally inexpensive, as it does not require any system modeling or involve any complex matrix operations;
4. Unlike traditional KF, LCF, and NCF, where attitude angles are considered as estimation states, here the error value is estimated using NCF, and then attitude angles are computed using LCF.

The feasibility of the proposed cascaded architecture is verified by comparing its results with the reference attitude parameters obtained from commercially available Attitude Heading and Reference System (AHRS) modules. As the inertial sensors contain time-varying bias, longer duration data are also used to validate the efficacy of the proposed framework. The obtained results are compared with the existing KF- and CF-based fusion algorithms discussed in the literature. The remaining part of the paper is organized as follows: **Section 2** presents the theoretical and mathematical background for CF and CKF, required for attitude estimation. The proposed cascaded complementary filter structure is presented in **Section 3**. The analysis of the proposed algorithm and its benchmarking with other algorithms are presented in **Section 4**. **Section 5** concludes the paper and provides the future direction of this work.

## 2. Mathematical Preliminaries

The orientation of a moving object is the angle made by the vehicle's body frame with respect to the world reference frame. The three different rotation angles for the x, y and z− axes (roll, pitch, and yaw) are generally denoted using the Greek letters phi (ϕ), theta (θ) and psi (ψ), respectively. The angular velocities measured by a gyroscope in the body reference frame for the x, y, and z− axes are generally denoted as p, q, and r, respectively. The relation between the time derivatives of the Euler angles and gyroscope measurements in the body frame is represented as Equations (1)–(3) [31]:

$$\dot{\phi}_g = p + q \sin\phi \tan\theta + r \cos\phi \tan\theta \qquad (1)$$

$$\dot{\theta}_g = q \cos\phi - r \sin\theta \qquad (2)$$

$$\dot{\psi}_g = p + q \sin\phi \sec\theta + r \cos\phi \sec\theta \qquad (3)$$

The attitude angles of roll, pitch, and yaw can be obtained by integrating the time derivatives from Equation (1) to Equations (3), respectively. However, this integration process accumulates the integration error, causing a drift in estimation. Hence, the orientation computed from the accelerometer and magnetometer is essential for accurate attitude estimation. The roll (ϕ) and pitch (θ) angles from the accelerometer can be computed using Equations (4) and (5), respectively [32], as

$$\phi_a = \tan^{-1}\left(\frac{a_y}{a_z}\right) \quad (4)$$

$$\theta_a = \tan^{-1}\left(\frac{-a_x}{a_y \sin\phi + a_z \cos\phi}\right) \quad (5)$$

Here, ax, ay and az are the accelerometer measurements along the *x, y,* and *z–* axes, respectively. Similarly, if mx, my and mz are the magnetometer measurements along x, y and z axes, respectively, then the yaw (ψm) angle using magnetometer can be computed as

$$\psi_m = \tan^{-1}\left(\frac{m_z \sin\phi - m_y \cos\phi}{m_x \cos\theta + m_y \sin\theta \sin\phi + m_z \sin\theta \cos\phi}\right) \quad (6)$$

In order to overcome the errors and to take advantage of the complementary nature of motion characteristics, sensor fusion techniques are used to estimate accurate attitude. The following section discusses the theoretical details of the complementary filter and complementary Kalman filter in detail.

### 2.1. Complementary Filter

The complementary filter is a computationally inexpensive sensor fusion technique that consists of a low-pass and a high-pass filter. In this application of inertial-sensor-based attitude estimation, the gyroscope's dynamic motion characteristics are complementary to that of the accelerometer and magnetometer. The basic structure of CF shown in **Figure 1** consists of two inputs, x1 and x2, which are low- and high-frequency noise-corrupted versions of the signal *x*. The complementary filter output x^ is given in Equation (**7**).

$$\hat{x} = x_1 G(s) + x_2 \overline{G}(s) \quad (7)$$

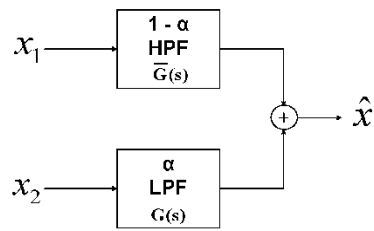

*Figure 1. Basic structure of Complementary Filter.*

Here, G(s) represents the transfer function for the low-pass filter, whereas $\overline{G}(s)$ is the transfer function of the high-pass filter, such that $G(s) + \overline{G}(s) = 1$.

Using this structure of CF for attitude estimation, gyroscope estimates, $\dot{x}_g(\dot{\phi}\dot{\theta}\dot{\psi})$, are applied at x1, and accelerometer/magnetometer estimates, $x_a(\phi_a\theta_a\psi_m)$ are applied at x2. Analytically, the attitude estimated using a linear CF structure, x^, is given as

$$\hat{x} = \alpha \left( \int \dot{x}_g dt \right) + (1 - \alpha)x_a \qquad (8)$$

The parameter $\alpha \in [0\,1]$ determines the weighing factor for gyroscope and accelerometer/magnetometer estimates. The LCF estimate, in terms of transfer function, can be represented as in Equation (9).

$$\hat{x} = \frac{\tau s}{1 + \tau s}\left(\frac{\dot{x}_g}{s}\right) + \frac{1}{1 + \tau s}(x_a) \qquad (9)$$

Here, the individual transfer functions of LPF and HPF can be represented as:

$$LCF\_HPF = \frac{\tau s}{1 + \tau s}; LCF\_LPF = \frac{1}{1 + \tau s}$$

Using these equations, the amplitude and phase plots are plotted together (**Figure 2**) to indicate the amplitude and phase plot of linear CF. It can be observed that the combined magnitude is unity (0 dB) and phase shift is 0 degrees over the complete frequency range.

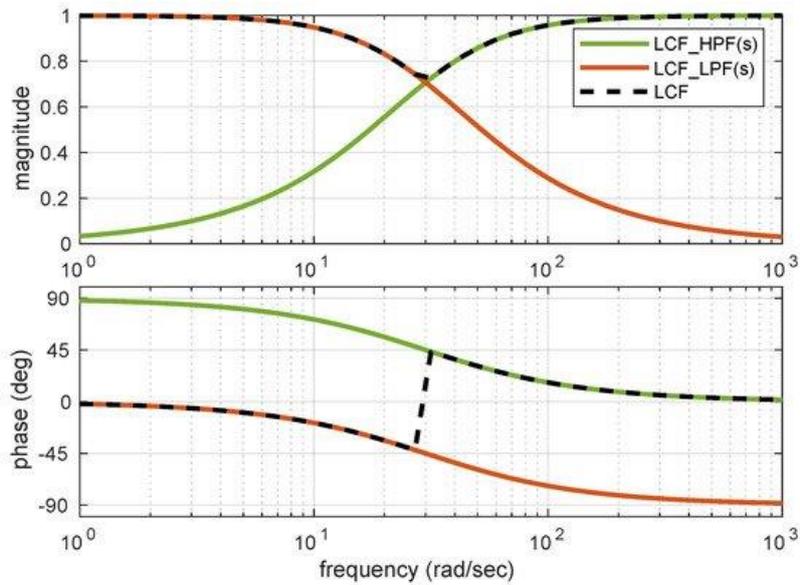

*Figure 2 Amplitude and Phase plot for LCF.*

However, this CF structure fails to estimate the gyroscope bias online and is unable to estimate the accurate attitude angles in the dynamic motion conditions [15].

The nonlinear complementary filter (NCF) shown in **Figure 3** uses the proportional-integral (PI) controller to reduce the steady-state error and compensate for the varying gyroscope bias. KP and KI indicates the proportional and integral gain, respectively, and the system estimate, x^ is represented as

$$\hat{x} = \frac{1}{s}\left[\dot{x}_g^e + \left(K_P + \frac{K_I}{s}\right)(x_a - \hat{x})\right] \quad (10)$$

$$\hat{x} = \frac{s^2}{s^2 + K_P s + K_I}\left(\frac{\dot{x}_g^e}{s}\right) + \frac{K_P s + K_I}{s^2 + K_P s + K_I}(x_a) \quad (11)$$

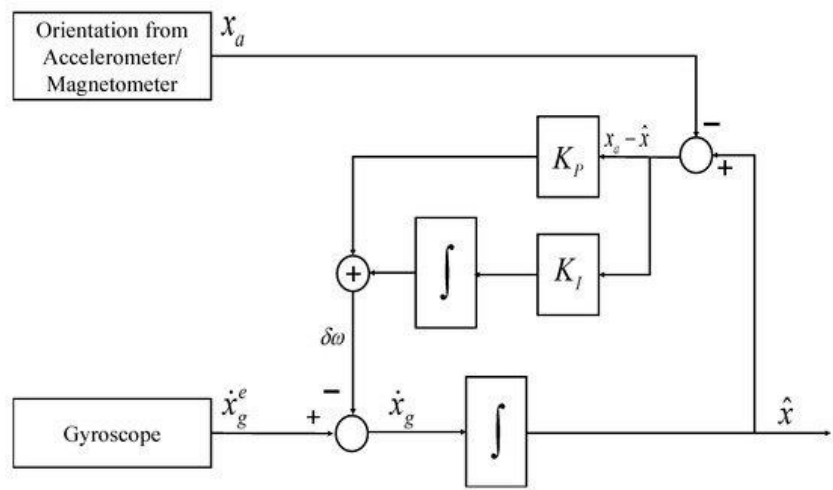

*Figure 3 Nonlinear structure of Complementary Filter.*

On similar lines of LCF, the estimate of the NCF is represented in terms of two transfer functions (Equation (11)), and amplitude and phase plots are plotted as shown in Figure 4. Unity magnitude and zero phase shift are observable over the complete frequency range.

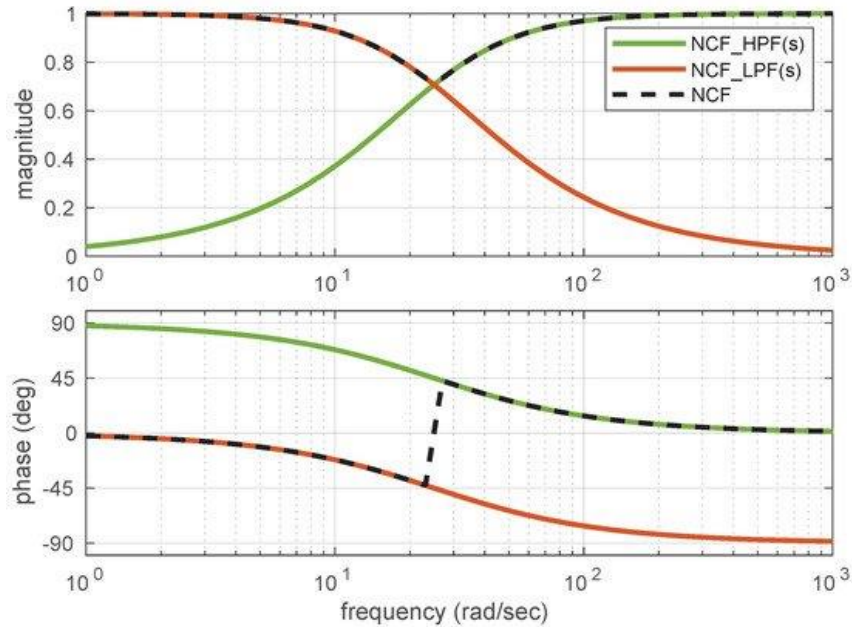

Figure 4 Amplitude and Phase plot for NCF.

Although CF is gaining popularity due to its simplicity, the Kalman filter is still a widely used technique for attitude estimation and has undergone several improvements. The KF technique comprises two steps, wherein the first step states are predicted using the process model, and, in the second step, the states are corrected using the measurement model. The Kalman filter works in a prediction–correction mechanism to yield a near-optimal state estimate under the assumption of Gaussian noise. Recent developments have shown that the filter performs well, even in the presence of colored noise [33]. The mathematical formulation of Extended Kalman Filter used in the attitude estimation is presented in Appendix A.

Another set of algorithms that takes advantage of both the complementary filter and the Kalman filter architecture has been proposed in the literature and is referred as Complementary Kalman Filter (CKF) [26,34], discussed in the following subsection.

### 2.2. Complementary Kalman Filter

Complementary Kalman Filter is a combination of Kalman filter and complementary filter in a single framework. In CKF, the Kalman filter is used for gyroscope error compensation, and the CF is then used for attitude computation. This provides the advantage of a rapid dynamic response from the system. In CKF, the error in measurements is generally considered as the system state for the KF, and the gyroscope bias is estimated. Unlike KF, the integration of the gyro rates is performed outside the KF block, in the "attitude computation" block. **Figure 5** shows the block-level structure of CKF. A detailed description of the CKF can be found in [26]. Although the KF and CKF are robust state estimators, they involve a large number of complex matrix-inverse operations, limiting their applications for low-cost systems. Additionally, the KF model needs to be appropriately formulated, considering the noise characteristics, before it is employed in any application. Contrary to KF, CF is a simple structured formulation that does not involve any complex mathematical operations. It does not consider any prior knowledge about noise characteristics or require system modeling.

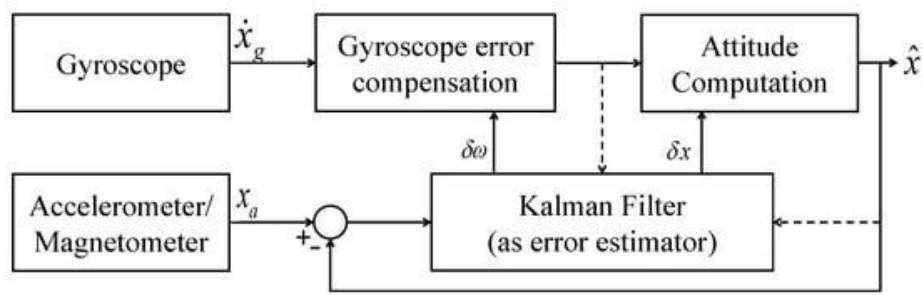

*Figure 5Complementary Kalman Filter.*

This paper proposes a two-stage complementary filter for attitude estimation. The NCF structure is used for the gyroscope error compensation, and a simple structured LCF is applied for attitude computation. The detailed, proposed gyro-error-compensated, attitude-estimation methodology is discussed in the following section.

## 3. Proposed Methodology

The proposed two-stage complementary filter, hereafter termed Cascaded Complementary Filter (CCF), is a combination of linear and non-linear versions of CF (NCF). The PI-controller-based NCF estimates the gyroscope bias, which is used to correct the gyroscope estimates. The corrected gyroscope measurements are then fused with the accelerometer/magnetometer measurement using linear CF. The proposed architecture of the cascaded complementary filter is shown in **Figure 6**. Here, $\dot{x}_{eg}$ represent the angular rates obtained through gyroscope measurements, $x_a$ represent the attitude computed using accelerometer/magnetometer readings, and $\alpha$, $K_P$ and $K_I$ are the filter constants.

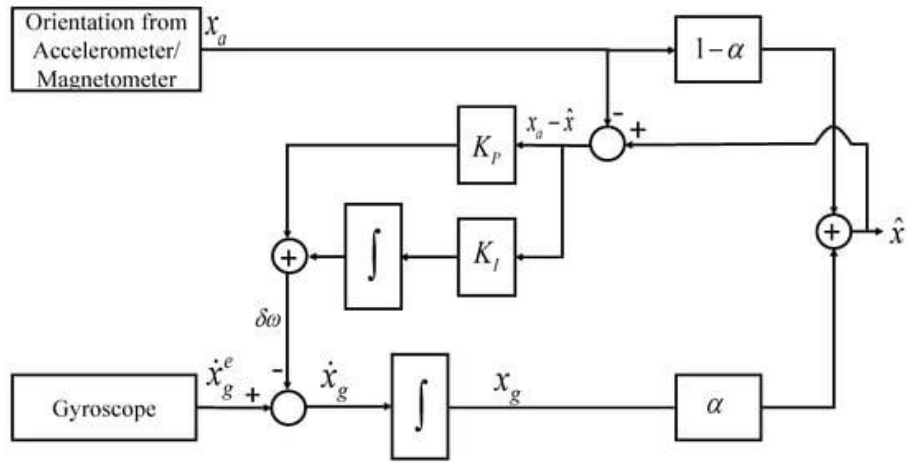

*Figure 6Cascaded Complementary Filter.*

The error value between the CCF-estimated attitude x^ and accelerometer/magnetometer-based attitude xa is used in gyroscope error compensation. The gyroscope error (δω) can be represented as

$$\delta\omega = \left(K_P + \frac{K_I}{s}\right)(x_a - \hat{x}) \qquad (12)$$

Using the final value theorem, Hong, in [35], proved that the error in estimated attitude angle (δx^) converges to error in the attitude angle estimated using an accelerometer, that is, limtime→∞δx^ = δxa. Hence, the error between attitude estimated from CCF and attitude computed from the accelerometer/magnetometer is applied to the PI controller for gyroscope-bias error computation. This error value is then subtracted from the angular rates obtained using a gyroscope to obtain the error-compensated gyroscope measurements given by Equation (13).

$$\dot{x}_g = \dot{x}_g^e - \delta\omega \qquad (13)$$

These angular rates are then integrated to obtain the attitude angles xg = (ϕg,θg,ψg). In the proposed architecture of CCF, the linear complementary filter is further used to combine the attitude angles from the gyroscope and those computed from the accelerometer/magnetometer. The attitude parameters x^ = (ϕ^,θ^,ψ^) obtained using the proposed CCF architecture can be mathematically represented as

$$\hat{x} = \alpha\left(\frac{1}{s}\left[\dot{x}_g^e + \left(K_P + \frac{K_I}{s}\right)(x_a - \hat{x})\right]\right) + (1-\alpha)(x_a) \qquad (14)$$

**Figure 7** represents the block diagram of the usage of linear and non-linear CF in the cascaded complementary filter structure. Although the linear and non-linear versions of CF are widely applied for attitude estimation, their primary disadvantage is the need for tuning filter parameters.

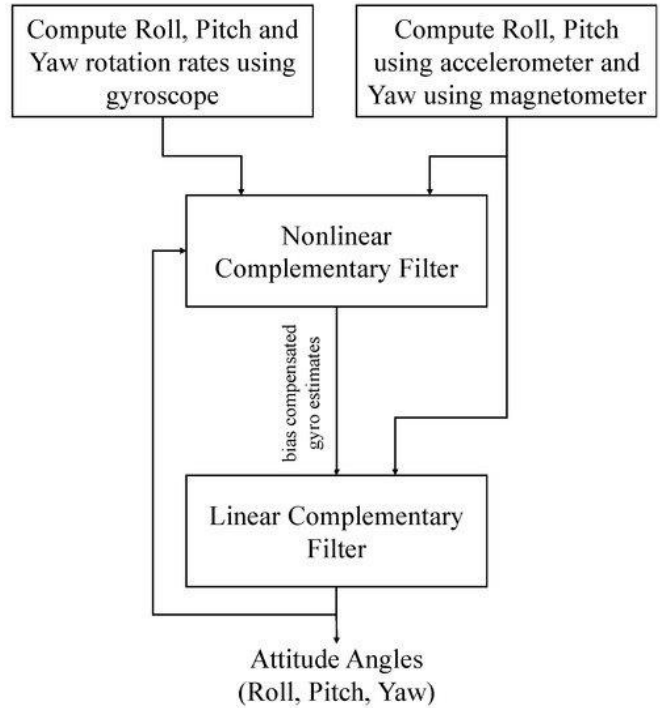

*Figure 7Flowchart of the proposed CCF algorithm.*

The CCF estimate Equation (14) can be rearranged in the form of two transfer functions (of LPF and HPF) as

$$\hat{x} = \frac{\alpha s^2}{s^2 + \alpha K_P s + \alpha K_I} \left(\frac{\dot{x}_g^e}{s}\right) + \frac{(1-\alpha)s^2 + \alpha K_P s + \alpha K_I}{s^2 + \alpha K_P s + \alpha K_I}(x_a) \quad (15)$$

Using these equations, the combined plots showing the amplitude and frequency responses are plotted and are shown in **Figure 8**. The amplitude and plase plots shown for LCF, NCF and CCF clearly indicate the all-pass filtering nature of these filters. The low-pass filtering of the accelerometer measurements and high-pass filtering of the gyroscope measurements together help to estimate the attitude angles in all the motion frequencies.

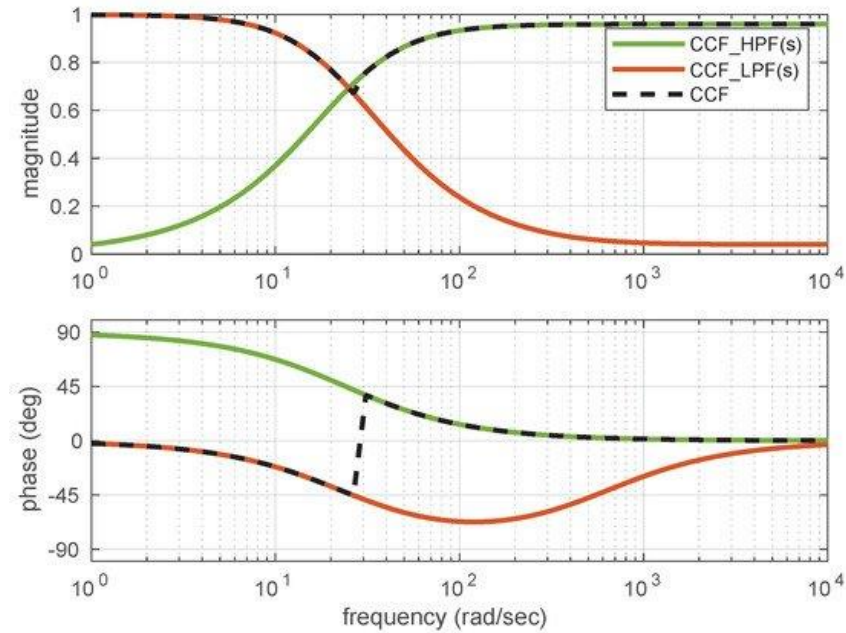

*Figure 8 Amplitude and Phase plot for CCF.*

Several techniques, such as fuzzy logic, neural networks, and optimization techniques, have been used in the literature for tuning filter parameters. The proposed CCF algorithm compensates for the gyroscope bias using a non-linear complementary filter, and then uses these bias-compensated gyroscope measurements in a linear complementary filter for attitude estimation. Contrary to the NCF, accelerometer/magnetometer measurements are used for gyroscope bias error correction, as well as for attitude angle computation. In the proposed architecture, the gyroscope error is compensated using NCF and the attitude angle is computed using LCF. The following section provides experimental proof, and the validation of the proposed CCF architecture as compared to other existing algorithms.

4. Results and Discussion

The cascaded complementary filter algorithm is applied to different datasets collected from commercially available AHRS modules. Relatively accurate Xsens MTI-G and comparatively cheaper Arducopter's APM navigation modules are used for data logging and benchmarking. The Arducopter module estimates the attitude angles based on CF [36], while the Xsens module has an internal KF for attitude estimation [37]. These modules have been given random motions independently in different directions to generate datasets and offline analysis. The data logged using these modules consist of raw sensor measurements along with their estimated attitude angles. Since these modules are commercially available, their estimates are obtained from the algorithm embedded in those platforms and are considered here as the reference attitude for comparison purposes. The raw sensor measurements are applied as an input to the proposed CCF algorithm, and the attitude angle computed through CCF is compared with the reference attitude angles. The root mean square error (RMSE) between the reference attitude angle xreference and the estimated attitude angle xmeasure is computed for quantitative comparison of the proposed algorithm, and is represented as

$$RMSE = \sqrt{\frac{1}{n}\sum_{k=0}^{n}(x_{reference}(k) - x_{measure}(k))^2} \qquad (16)$$

The overall experimentation is carried out in two broad categories. In the first category, the effect of varying CCF parameters is investigated and compared against the traditional non-linear complementary filter (NCF). In the second category, the proposed CCF results are benchmarked with other state-of-the-art schemes in terms of accuracy and computational complexity. These experiments are carried out using MATLAB 2020b, installed on a computer system with 4 GB RAM.

In the first category of investigations, the performance of CCF is analyzed by varying the gain parameters KP & KI, while maintaining a constant value of $\alpha=0.7$. The RMSE error values are compared with the error values obtained for NCF with the same values of KP and KI. **Table 1** and **Table 2** compare the average RMSEs obtained when KP and KI parameters are varied for a dataset captured through Xsens and Arducopter sensor modules, respectively. The average RMSE value in each row refers to the mean of the RMSEs obtained for ϕ, θ, and ψ. If RMSEϕ, RMSEθ and RMSEψ represents the RMSE obtained in roll, pitch and yaw, respectively, then the average RMSE (RMSEaverage) value is computed as

$$RMSE_{average} = \frac{RMSE_\phi + RMSE_\theta + RMSE_\psi}{3} \qquad (17)$$

The datasets logged using the Xsens module are referred to as X1, X2, X3, and X4, whereas those logged from the Arducopter module are referred to as A1, A2, A3, and A4.

Table 1 Average RMSE (in radian) obtained for NCF and CCF with varying $K_P$ and $K_I$ (Xsens datasets).

| Xsens Dataset | | X1 | | X2 | | X3 | | X4 | |
| --- | --- | --- | --- | --- | --- | --- | --- | --- | --- |
| $K_p$ | $K_I$ | NCF | CCF | NCF | CCF | NCF | CCF | NCF | CCF |
| 75 | 0.01 | 0.042 | 0.041 | 0.077 | 0.077 | 0.076 | 0.075 | 0.077 | 0.077 |
| 75 | 0.1 | 0.042 | 0.041 | 0.077 | 0.077 | 0.076 | 0.075 | 0.077 | 0.077 |
| 75 | 1 | 0.042 | 0.041 | 0.077 | 0.077 | 0.076 | 0.076 | 0.077 | 0.077 |
| 25 | 0.01 | 0.046 | 0.041 | 0.076 | 0.077 | 0.077 | 0.075 | 0.075 | 0.076 |
| 25 | 0.1 | 0.046 | 0.041 | 0.076 | 0.077 | 0.077 | 0.075 | 0.075 | 0.076 |
| 25 | 1 | 0.046 | 0.041 | 0.076 | 0.077 | 0.078 | 0.075 | 0.075 | 0.076 |
| 1 | 0.01 | 0.264 | 0.040 | 0.199 | 0.076 | 0.248 | 0.075 | 0.324 | 0.075 |
| 1 | 0.1 | 0.276 | 0.040 | 0.200 | 0.076 | 0.262 | 0.075 | 0.338 | 0.075 |
| 1 | 1 | 0.383 | 0.040 | 0.210 | 0.076 | 0.307 | 0.075 | 0.404 | 0.075 |
| 0.1 | 0.01 | 0.436 | 0.040 | 2.303 | 0.076 | 0.470 | 0.075 | 0.848 | 0.075 |
| 0.1 | 0.1 | 1.237 | 0.040 | 1.293 | 0.076 | 0.753 | 0.075 | 1.264 | 0.075 |
| 0.1 | 1 | 9.118 | 0.040 | 3.665 | 0.076 | 2.038 | 0.075 | 6.566 | 0.075 |
| mean RMSE | | 0.998 | 0.040 | 0.694 | 0.077 | 0.378 | 0.075 | 0.850 | 0.076 |
| standard deviation | | 2.470 | 0.000 | 1.110 | 0.001 | 0.539 | 0.000 | 1.760 | 0.001 |
| LSE adaptive | | 0.043 | 0.041 | 0.077 | 0.077 | 0.076 | 0.076 | 0.076 | 0.076 |

Table 2 Average RMSE (in radian) obtained for NCF and CCF with varying Kp and KI (Arducopter datasets).

| Arducopter Dataset | | A1 | | A2 | | A3 | | A4 | |
| --- | --- | --- | --- | --- | --- | --- | --- | --- | --- |
| $K_p$ | $K_I$ | NCF | CCF | NCF | CCF | NCF | CCF | NCF | CCF |
| 75 | 0.01 | 0.269 | 0.268 | 0.447 | 0.418 | 0.090 | 0.088 | 0.208 | 0.194 |
| 75 | 0.1 | 0.269 | 0.268 | 0.447 | 0.418 | 0.090 | 0.088 | 0.208 | 0.194 |
| 75 | 1 | 0.269 | 0.268 | 0.450 | 0.418 | 0.090 | 0.088 | 0.208 | 0.194 |
| 25 | 0.01 | 0.322 | 0.280 | 0.490 | 0.409 | 0.098 | 0.075 | 0.248 | 0.204 |
| 25 | 0.1 | 0.322 | 0.280 | 0.490 | 0.409 | 0.098 | 0.075 | 0.247 | 0.203 |
| 25 | 1 | 0.322 | 0.280 | 0.495 | 0.409 | 0.098 | 0.075 | 0.245 | 0.203 |
| 1 | 0.01 | 0.913 | 0.301 | 2.122 | 0.421 | 0.541 | 0.073 | 2.481 | 0.208 |
| 1 | 0.1 | 0.936 | 0.301 | 3.325 | 0.421 | 0.521 | 0.073 | 3.839 | 0.208 |
| 1 | 1 | 1.032 | 0.301 | 8.585 | 0.422 | 0.429 | 0.073 | 2.675 | 0.211 |
| 0.1 | 0.01 | 2.871 | 0.302 | 7.463 | 0.422 | 3.668 | 0.073 | 6.873 | 0.211 |
| 0.1 | 0.1 | 1.344 | 0.302 | 16.402 | 0.422 | 3.966 | 0.073 | 64.488 | 0.211 |
| 0.1 | 1 | 1.598 | 0.302 | 23.814 | 0.423 | 5.559 | 0.073 | 12.941 | 0.211 |
| mean RMSE | | 0.872 | 0.288 | 5.377 | 0.418 | 1.270 | 0.077 | 7.888 | 0.204 |
| standard deviation | | 0.752 | 0.014 | 7.277 | 0.005 | 1.860 | 0.006 | 17.453 | 0.007 |
| LSE adaptive | | 0.292 | 0.288 | 0.451 | 0.423 | 0.083 | 0.080 | 0.221 | 0.219 |

**Table 1** and **Table 2** present the NCF and CCF performance on eight different datasets, while the gain parameters are varied. It can be observed that, for NCF, the error values vary significantly with different combinations of KP and KI parameters, whereas, for CCF, these error values are almost constant. The average RMSE value for CCF is approximately equal to its mean value and is also approximately equal to the RMSE value obtained while NCF is tuned adaptively with the LSE-aided NCF (LSCF) [38]. However, in the case of NCF, the RMSEs vary drastically from their mean value as well as from the RMSEs obtained from LSCF. It was, therefore, concluded that the variation in KP and KI affects the NCF performance significantly, whereas the proposed CCF framework is independent of this. A nearly zero standard deviation in the case of CCF, and a non-zero standard deviation for NCF, further supports the claims regarding CCF. Our previous work, presented in [38], aims at the adaptive tuning of the NCF using the Least Square Estimation (LSE) technique. In the current research work, an attempt is also made to tune the parameters of CCF, i.e., KP, KI, and $\alpha$, using the LSE technique. The obtained values for these parameters using LSE techniques are shown in Table 3. The obtained RMSEs for the LSE-tuned CCF are mentioned in the last rows of Table 1 and Table 2. It can be observed that the average RMSE values for CCF without adaptation are similar to those obtained with adaptation for all the datasets. This depicts that the inclusion of LSE does not provide any additional advantage, and the change in CCF filter parameters is inessential. It can also be noted that the mean RMSE obtained for CCF is approximately equal to the RMSE obtained when NCF is adaptively tuned using the LSE technique [38].

*Table 3 Values of LSE-adapted parameters of CCF.*

| Dataset | X1 | X2 | X3 | X4 | A1 | A2 | A3 | A4 |
|---|---|---|---|---|---|---|---|---|
| **Roll** | | | | | | | | |
| $K_P$ | 21.170 | 38.336 | 33.743 | 6.746 | 21.118 | 30.309 | 23.267 | 39.408 |
| $K_I$ | 5.131 | 5.003 | 7.047 | 0.431 | 4.038 | 5.174 | 4.631 | 4.356 |
| $\alpha$ | 0.743 | 0.934 | 0.884 | 0.750 | 0.643 | 0.777 | 0.885 | 0.951 |
| **Pitch** | | | | | | | | |
| $K_P$ | 2.807 | 35.431 | 38.290 | 39.574 | 1.229 | 34.542 | 39.940 | 36.182 |
| $K_I$ | 0.348 | 6.200 | 5.198 | 3.755 | 0.301 | 6.255 | 4.225 | 6.513 |
| $\alpha$ | 0.419 | 0.902 | 0.941 | 0.959 | 0.667 | 0.858 | 0.955 | 0.888 |
| **Yaw** | | | | | | | | |
| $K_P$ | 39.672 | 42.463 | 21.345 | 38.266 | 42.334 | 42.544 | 40.241 | 42.094 |
| $K_I$ | 4.164 | 0.569 | 4.885 | 4.896 | 0.719 | 0.460 | 4.733 | 0.970 |
| $\alpha$ | 0.960 | 0.995 | 0.709 | 0.932 | 0.993 | 0.996 | 0.974 | 0.989 |

Further analyses were carried out to investigate the effect of varying the $\alpha$ parameter while the KP and KI parameters are kept constant. Figure 9 plots the RMSE values for two different combinations of KP and KI on the A3 dataset, while the $\alpha$ value is varied from 0.1 to 0.9. This graphical comparison for the average RMSEs proves the independence of CCF performance from the value of the $\alpha$ parameter.

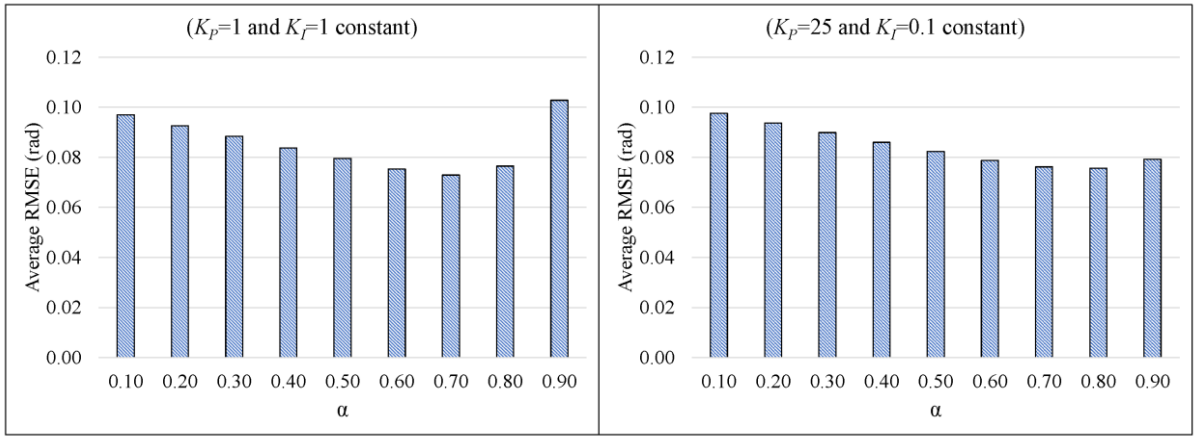

*Figure 9 Effect of varying α value on the CCF performance with fixed KP and KI parameters (Dataset A3).*

It can be noted from **Table 1** and **Table 2** and **Figure 9** that the variation in the computed RMSE is negligible with the parameters' variation in the case of the proposed CCF architecture. In the case of LCF, the estimation accuracy is highly affected, as there is no provision for the computation of gyroscope bias in runtime. In the case of CCF, firstly, the gyroscope bias error is compensated using the NCF structure, and then the attitude angles are estimated using the LCF structure. Further experimentation has been carried out to compare the accuracy of CCF estimation results with respect to the other existing state-of-the-art attitude estimation algorithms.

In the second category of experimentation, the proposed CCF algorithm is benchmarked against other state-estimation algorithms. The attitude estimated using the CCF scheme is compared with a Kalman-filter-based cascaded structure (CKF) and the Extended Kalman Filter (EKF), LSE-aided adaptive NCF (LSCF), traditional non-linear complementary filter (NCF), and popular non-linear complementary filter algorithms, namely, the Mahony filter and Madgwick filter. In these comparisons, the parameter values chosen for investigation were KP = 25, KI = 0.1 and $\alpha$ = 0.7.

**Figure 10** and **Figure 11** show the comparison of RMSEs of $\phi$, $\theta$ and $\psi$ of the proposed CCF structure with other estimation algorithms on Xsens and Arducopter datasets, respectively. In both these figures, the x-axis represents different datasets, and the y-axis represents the average RMSE value in radians. It can be observed that the RMSE values obtained using the CCF scheme are lesser than/comparable to the other techniques. In this experimentation, the filter parameters of NCF, CKF, and EKF were selected based on the trial-and-error method. The parameters for the Mahony and Madgwick filters were also tuned manually. **Table 4** indicates the different parameter values used in this paper for simulation. The MATLAB implementation for Mahony and Madgwick filter algorithms was obtained from [39]. It is nontable that the LSCF [38] was the adaptive algorithm used to tune the filter parameters. Even though the parameters for the CCF algorithm were chosen randomly, its performance was almost equivalent to the adaptive algorithm.

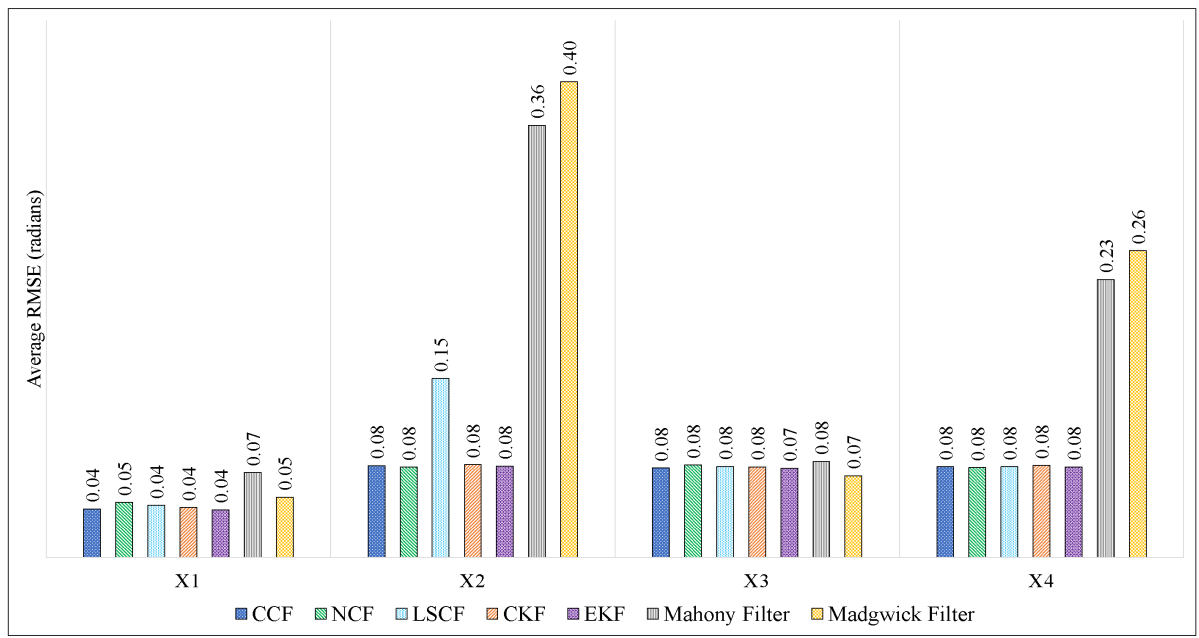

Figure 10 Comparison of proposed CCF structure with other estimation algorithms on Xsense dataset.

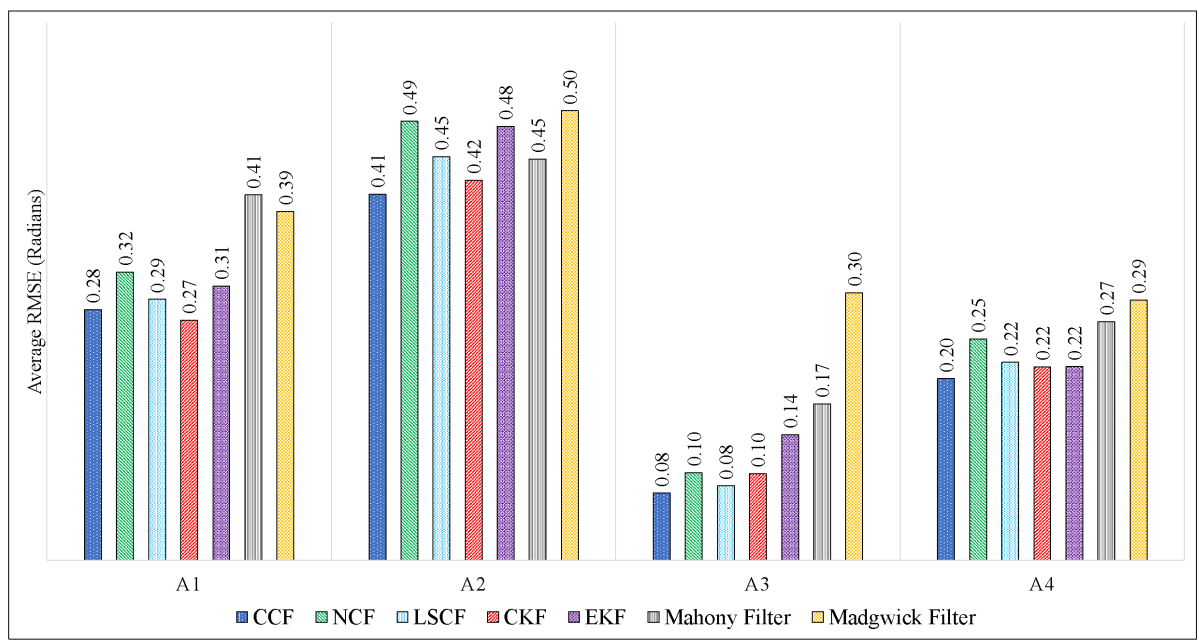

Figure 11 Comparison of proposed CCF structure with other estimation algorithms on Arducopter datasets.

Table 4 Considered parameter values for different algorithms.

| | Filter Parameter | Xsens Dataset | Arducopter Dataset |
|---|---|---|---|
| Mahony Filter | $K_P$ | 1.5 | 100 |
| Madgwick Filter | $\beta$ | 0.2 | 10 |
| NCF | $K_P$ | 25 | 25 |
| | $K_I$ | 0.1 | 0.1 |
| CKF | $Q$ | | $\begin{bmatrix} 60 & 0 & 0 & 0 & 0 & 0 \\ 0 & 5 & 0 & 0 & 0 & 0 \\ 0 & 0 & 10 & 0 & 0 & 0 \\ 0 & 0 & 0 & 0 & 0 & 0 \\ 0 & 0 & 0 & 0 & 0 & 0 \\ 0 & 0 & 0 & 0 & 0 & 0 \end{bmatrix}$ |
| | $R$ | | $\begin{bmatrix} 0.1 & 0 & 0 \\ 0 & 0.01 & 0 \\ 0 & 0 & 0.8 \end{bmatrix}$ |
| EKF | $Q$ | | $\begin{bmatrix} 5 & 0 & 0 \\ 0 & 0.05 & 0 \\ 0 & 0 & 10 \end{bmatrix}$ |
| | $R$ | | $\begin{bmatrix} 0.1 & 0 & 0 \\ 0 & 0.001 & 0 \\ 0 & 0 & 0.8 \end{bmatrix}$ |

The computation time for the CKF and EKF algorithms was much higher than that of the CCF algorithm, which is due to the complex matrix operations involved in the KF structure. **Figure 12** compares the relative time required by CCF, CKF, and EKF algorithms on different datasets. In this comparison, the algorithm was simulated on the same dataset 20 times, and the average simulation time was considered for comparison. The simulation time depends on the computer system parameters on which the simulations are carried out. Normalized simulation time is shown in **Figure 12** for appropriate comparison.

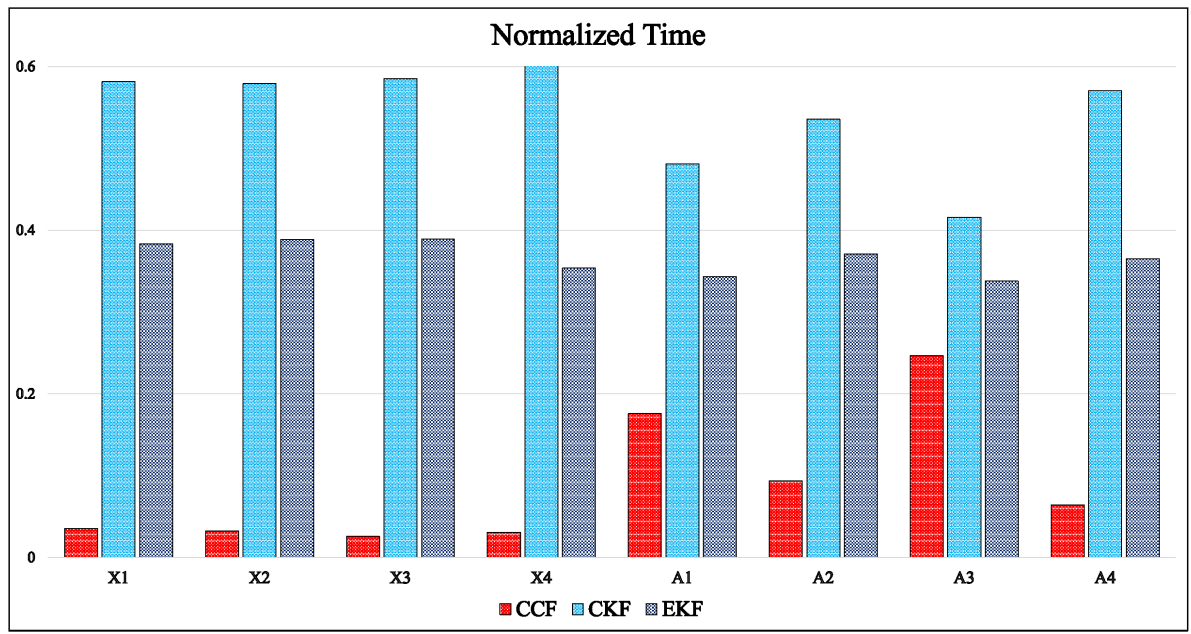

Figure 12 Comparison of computational time for Xsens and Arducopter datasets.

For purposes of brevity, the attitude estimation results obtained using the CCF scheme are shown in **Figure 13** and **Figure 14** for Xsens and Arducopter datasets, respectively. These figures compare the proposed CCF algorithm to the attitude angles obtained using the gyroscope and accelerometer/magnetometer alone for the logged dataset. In these figures, red curves represent the reference attitude estimates from the AHRS modules. The attitude computed using gyroscope alone is represented using green dotted curves, whereas blue-color lines represent the attitude computed using an accelerometer/magnetometer alone. The attitude estimated using the proposed CCF architecture is represented using the black dashed curves. The estimates through EKF and NCF are also shown for reference purposes.

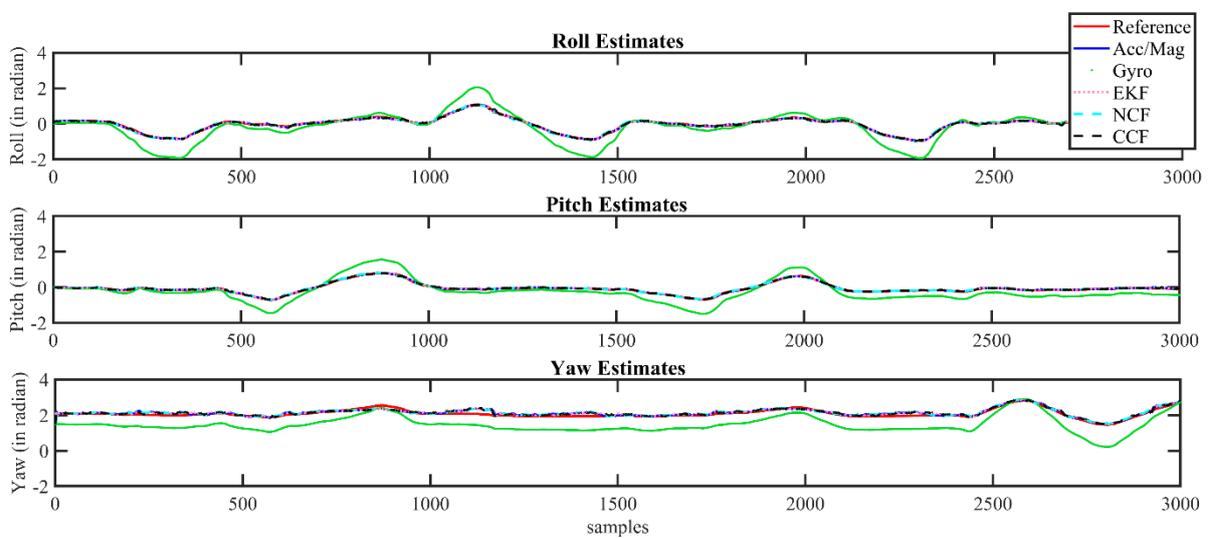

*Figure 13 Attitude estimation results for Xsens dataset.*

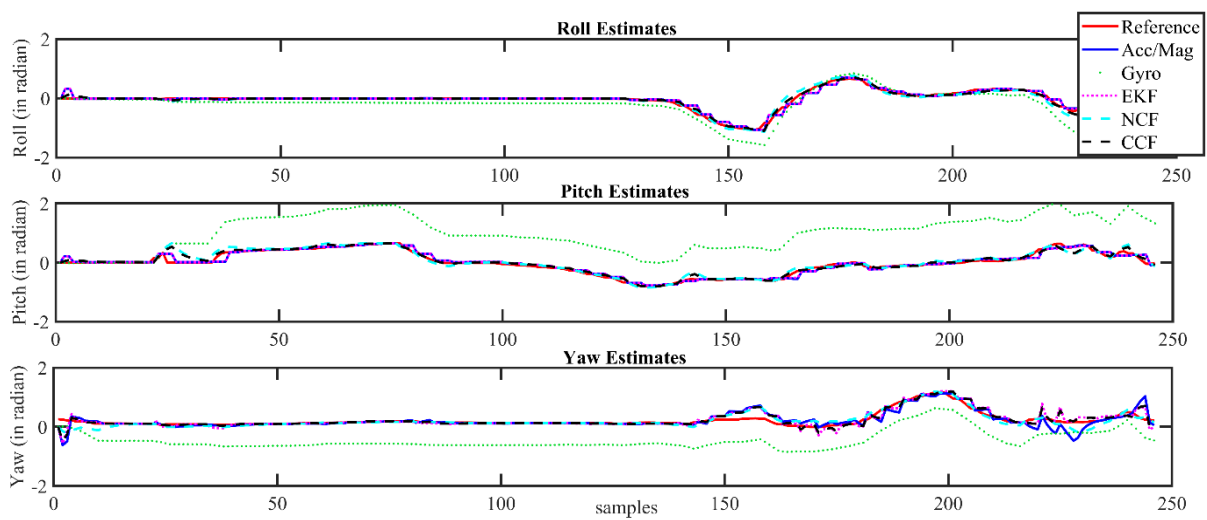

*Figure 14 Attitude estimation results for Arducopter dataset.*

It can be observed that the attitude estimated using the gyroscope alone starts to drift after a certain time interval. The estimates from the accelerometer/magnetometer contain the flicker noise. However, the proposed sensor-fusion-based CCF algorithm overcome these issues and precisely tracked the reference trajectory obtained from the commercial AHRS modules. It can be observed that the attitude computed using a gyroscope alone started to drift after a certain time interval, while that from the accelerometer/magnetometer contained flicker noise. A sensor fusion architecture could overcome these individual sensor issues and help to obtain a reliable estimate. In this case, the estimation results of CCF during the initial period of motion (specifically in the case of Arducopter dataset) is not as good as the attitude computed using the accelerometer alone; it cannot be generalized for the complete sequence. During initial time instants, the sensor bias is small and grows over time, while the fusion algorithms take some time to compensate the errors, thereby modifying the weights for different sensor outputs. The results show that the sensor fusion outperforms the individual sensor estimates.

Inertial sensors have time-varying bias characteristics, which also drift over time. Hence, further experimentation was carried out to investigate the performance of the proposed CCF over a dataset of longer-range duration. Simulated data were generated for almost 2 h using a sensor fusion and tracking toolbox, available in MatLab. Sensors were modeled, and random time-varying noise was added to sensor measurements. The proposed CCF was applied to this longer-duration dataset, and the results are presented in **Figure 15**. The Figure also shows the estimation results obtained through EKF, CKF and NCF. The red curves represent the reference/ideal attitude angles generated through Matlab; the EKF estimation is shown in blue, NCF estimates are indicated using a dotted green line and the estimation of CCF is shown with black dashed lines. It can be observed that, even though the measurements have significant noise, the proposed framework can compensate for the noise and provide accurate estimations. To indicate the estimation error over the time, RMSE is computed for every 1000 s and plotted in **Figure 16** for indication and reference purposes. The results also depict that the proposed CCF algorithm provides a feasible solution to the attitude estimation, without incorporating any complex adaptive-tuning algorithm. Through all the experimentation carried out, it is observed that, although the proposed CCF does not show any improvement in accuracy compared to the existing algorithms, it is computationally fast. Additionally, the structure does not require any tuning parameter and is reliable alternative to attitude estimation tasks for low-cost applications. The main contribution of this paper is its arrival at a filter with minimal or no tuning parameters, unlike other filters, whose performance largely depends on different filter parameters.

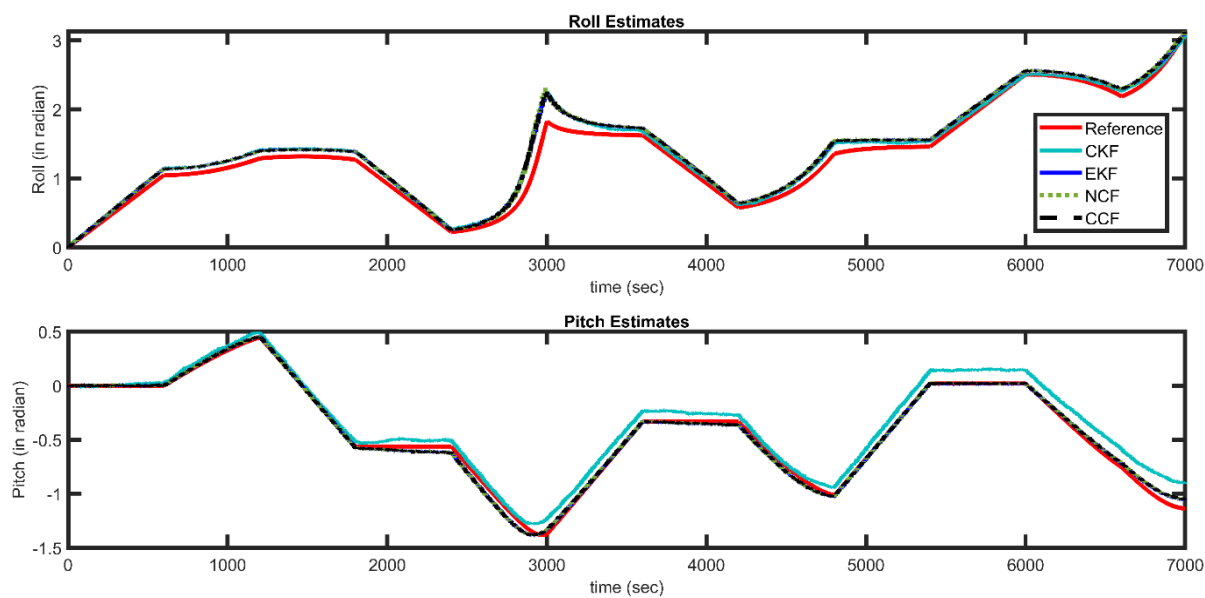

Figure 15 Attitude estimation results Matlab generated dataset.

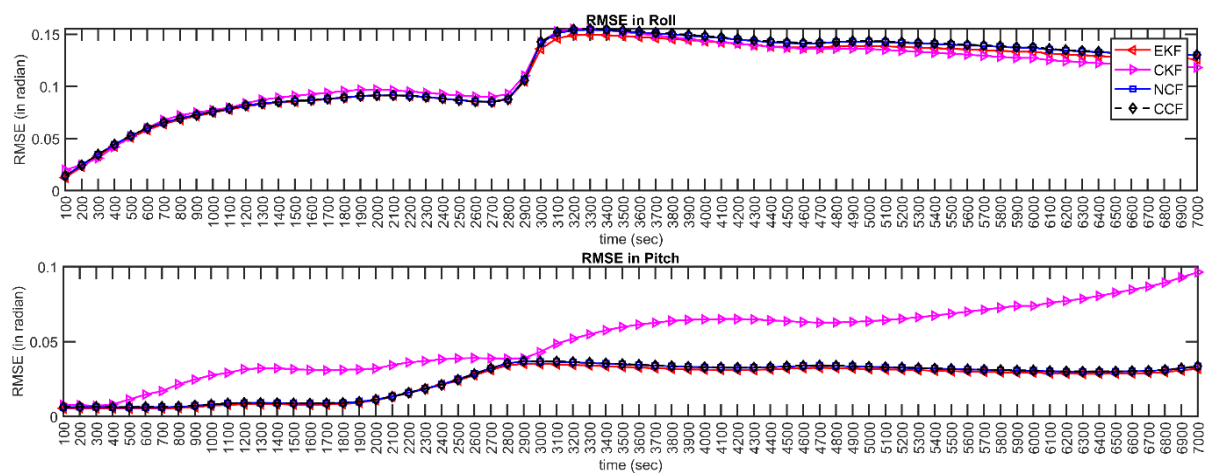

Figure 16 Estimation error over time: RMSE for every 100 s.

5. Conclusions

The paper presents a novel, cascaded, complementary, filter-based sensor fusion for attitude estimation applications. The system considers data obtained from an accelerometer, gyroscope, and magnetometer for sensor fusion. The proposed structure cascades the linear and non-linear version of a complementary filter and is inspired by the cascaded Kalman filter architectures. The proportional integral-based, non-linear version of CF is used to compute the gyroscope bias online, and the linear version is used to estimate the attitude parameters. The proposed architecture does not require any tuning (manual or adaptive) for the selection of filter parameters and is computationally inexpensive. The CCF technique has been compared with other existing algorithms and an adaptive variant of complementary filters to prove its efficacy. It is found that this scheme has a similar accuracy to the other schemes, with a very low deviation in changing gain parameters, demonstrating its success on different datasets. Even though the proposed framework does not provide an improvement in the estimation accuracy, it is a suitable alternative to attitude estimation, without any dependency on tuning filter parameters.

In future, it is planned to validate the algorithm using accurate rotary tables in a controlled environment. Accurate estimation of attitude angles is essential in the velocity and position estimation systems. In the future, it is also planned to extend the work to velocity and position estimation, where the CF structures can also be compared with full EKF frameworks. The work would also aim to explore artificial-intelligence-based reinforcement learning techniques for performing sensor fusion.